# Supporting Students with ADHD in Introductory Physics Courses: Four Steps for Instructors

Caroline Bustamante, Erin Scanlon, & Jacquelyn J. Chini

University of Central Florida

## 1. Introduction

The number of students with disabilities and specifically students with attention-deficit hyperactivity disorder (ADHD) entering postsecondary STEM education has been increasing in recent decades. However, many instructors and popular research-based curricula are not prepared to support such learner variation. The views and experiences of people with disabilities are not uniform, either across diagnoses or within a single diagnosis, such as ADHD. Thus, individuals' thoughts about how to support students with ADHD in physics courses will vary. We present views from one student with ADHD about strategies instructors can use to help her succeed in introductory physics courses.

## 2. Students with Disabilities in the Postsecondary Environment and STEM

Students with disabilities are enrolling in postsecondary education at increasing rates; while 10.5% of students enrolled in science and engineering degree programs in 2014 identified with disabilities, the percentage increased to 19.8% in 2017 (as measured using expanded definitions related to disability).[1-2] The prevalence of specific diagnoses with which students identify has also been changing. For example, the percentage of students with disabilities who identified with attention-deficit disorder (ADD)[3] more than tripled from 6.7% in 2000 to 21.8% in 2011.[4-5]

Multiple studies have found that faculty lack awareness of the legal requirements related to accommodations,[6-7] lack knowledge of inclusive pedagogies,[8] state that they do not feel prepared to teach students with disabilities,[9] and want more training related to accessibility.[10] Introductory research-based physics curricula have been shown to miss opportunities to enact strategies for supporting learner variation.[11] This combination of unprepared instructors and curricula not designed to support students with disabilities likely means these students are not well served in postsecondary physics courses.

It is essential that we address this concern through a combination of preparing faculty, creating more accessible and inclusive curricula, and increasing overall knowledge about disability in the physics community. It is also essential that this process starts from the experiences of students with disabilities,[12] while recognizing that each students' experience is unique. Without input from students with disabilities, instructors may add practices with the intention of increasing accessibility that may not benefit students or, even worse, may create additional barriers. For example, Harshman, Bretz, and Yeziersk (2013) found that accommodations developed by sighted faculty were overwhelming and therefore not useful for chemistry learners with visual impairments.[13] We present the viewpoint of a student diagnosed with attention-deficit hyperactivity disorder (ADHD)[14] about instructional practices that did or could have supported her learning in introductory physics courses.

Specifically, this student completed introductory physics 1 (mechanics) and 2 (electricity and magnetism) at a large southeastern research-intensive university in her third and fourth years in college, respectively. Both courses were SCALE-UP (Student-Centered Active Learning Environment with Upside-Down Pedagogies) style,[15] which meant they occurred in a classroom designed for student-centered group work and featured increased group work and decreased lecture compared to a traditional lecture-style course. She majored in biomedical sciences so she took multiple other STEM courses at the same institution. She was also registered with the university's Disability Services Office and received accommodations for both courses, including extended time on exams and quiet, secluded environment for taking exams.

3. Ways to Support Students with ADHD in Introductory Physics Courses

Physics instructors may encounter students who give them blank stares or confused looks but who are engaged in lecture. Or students who are repeatedly asking the same question as they are visibly taking notes. When the students take their exams, their scores are not reflective of their understanding of material. Their homework is sometimes late or incomplete. At times they are very social and may even disrupt class due to boredom or loss of focus. You may view them as "class clowns", "smart alecks", "trouble-makers", or "squeaky wheels". While instructors may come up with a myriad of explanations for these behaviors, it is important to be aware they can be associated with ADHD.

Students with ADHD may have difficulties and needs that differ from their peers and can hinder their performance in introductory physics courses. Students with ADHD can: 1) easily become distracted during class; 2) get lost in the fine details of a problem or bogged down with the minutia of each lesson; 3) get bored or distracted when class does not interest or challenge them; 4) become overwhelmed by processing physics content via multiple means at the same time (i.e., seeing it written on the board, hearing the instructor talking about it, and writing about it in their notes);[16] 5) require more time to process ideas than instructors believe; and 6) feel frustrated that they are not "getting it" when they see their peers excelling.

Thus, instructors need to be aware that some of the typical difficulties they see in their students may not be due to the commonly believed reasons (e.g., students lacking understanding or prerequisite knowledge, not being able to handle the course, not putting in the time and effort to learn the material) but instead may be related to a non-apparent disability such as ADHD. With this new understanding, instructors can tackle student difficulties using a different lens that calls for different instructional strategies. Below, we discuss four strategies instructors can use to help support students with ADHD in introductory physics courses. While these strategies center students with ADHD, they are likely to benefit all students by providing options and supports for variation in learners' needs, abilities, and interests.[17] Additionally, we believe that all of our suggestions are generally applicable beyond introductory physics courses. However, the second and third suggestions are particularly important in introductory physics courses because of the emphasis on conceptual learning over rote memorization in physics courses[18] which can be more taxing to students with ADHD. Therefore, having course material available inside and outside of class (third suggestion) in an organized fashion (second suggestion), can help some students with ADHD be successful in introductory physics courses. 3.1 Instructors Should Initiate Open Dialogue about Students' Needs, Abilities, and Interests

An approachable instructional demeanor inside and outside of class can help all students feel welcome. During the first few days of class, students are getting to know their instructors and may be assessing how amenable the instructor is to hearing and addressing their concerns. Students with ADHD will be gauging the instructor on approachability, respect towards students with different learning needs, willingness to provide options for learning, and confidentiality.[19] Students with disabilities can find it daunting to talk about their disabilities with a new instructor.[20] Research has indicated that the success of students with disabilities in postsecondary education is related to faculty attitudes toward disability.[21-22] Open dialog is necessary to identify supports and options that maintain the rigor of the course while optimizing learning for the particular student.

Instructors can demonstrate that they are approachable to students with disabilities by using class time to discuss the federally mandated accommodation[23-24] statements likely present on their syllabus. Many instructors only give a cursory description of these statements or skip them entirely; this can make students with disabilities feel as though the instructor does not actually care about them or their learning. Instructors should describe how students can contact them to discuss and arrange the accommodations and the supports they need. This initial presentation and discussion sets the tone for interactions between the student and instructor throughout the semester.

Another way to create an open classroom environment is to normalize student errors by pointing out common mistakes that students make in the course. This can be done by telling students that errors are natural and useful to the process of learning. Students with disabilities are routinely told that they are not "up to par" or negatively compared with their peers solely due to their diagnosis. Your disposition as a friendly instructor can alleviate a lot of the students' anxieties about succeeding in the physics course.

3.2 Scaffold the Course Content to Help Students Stay on Track

Students with ADHD can have difficulties focusing, especially during class time. This can lead them to lose the flow of the lecture, thereby missing some of the content covered. For example, if the instructor is drawing a free body diagram, a student with ADHD could have the following train of thought:

> *What are the forces at work? How I do to set up equations? What goes on the left-hand side? What goes on the right-hand side? How do I set them equal to each other? What equation do I need? Tension? Weight? Weight equals m times g. But what about the initial velocity? Do I need the initial velocity? Do I need the final? Which do I need? If I visualize myself as the object in motion – I am the sled on a* *ramp, which forces are acting against me. What is relevant in this case? What is required in my equation? How do I get velocity? What don't I need? what doesn't apply…*

And the thoughts can go on and on. While this thought process represents questions instructors want students to consider during problem solving, engaging in this internal dialogue interferes with students' attention to the on-going instructor demonstration of the solution, causing the student to miss the demonstration and possibly miss the transition to the next topic.

We suggest that instructors carefully scaffold their courses and communicate in multiple forms the scaffolds to help students with ADHD. For example, instructors should post a course schedule in the syllabus or on the learning management system (LMS), provide a concept map for each book chapter, and state the learning objectives for each class period.[25-26] Such scaffolds allow students with to stay on track and to get them back on track if they get distracted. Scaffolding can tell students what to expect during class, how to prepare for class and exams, and can allow students to see how each lesson relates to a main topic and the connections between main topics.

3.3 Provide Course Resources in Multiple Formats to Allow for Options in How and When Students Learn Content

Some students with ADHD need more clarification on problems than other students. For example, some students with ADHD need to see multiple, similar problems worked out step-by-step to help solidify the process in their memory and notice variations to the process across individual problems. Because many students with ADHD get distracted during class, repeating the problem-solving process can help fill in the information they may have missed or not fully understood the first time. Students with ADHD also need practice in finding the mental pathways to follow as they are solving problems (i.e., build up their problem-solving decision tree in their minds).

Providing resources to students outside of class in multiple formats (e.g., detailed, written problem solutions, videos of instructor solving problems) allows students to engage with course material as many times as they would like, thereby providing students options for when and how to engage with course material outside of class.[27] Instructors can also post their well-organized lecture slides to the LMS, encourage participation in office hours and provide students with resources that present course material in a different format (e.g., providing YouTube videos on a topic that is also covered in the textbook). Allowing students multiple means through which to access the information allows students to work with the mode that is most helpful for them. For students with ADHD, this is particularly important because: 1) working with multiple means can allow them to keep focused longer, 2) it allows them to work through a myriad of problems to build up their mental decision trees, and 3) it allows them to cycle through the material so that if they get distracted or lost they can continue to learn on their own without continually requiring the assistance of an instructor.

3.4 Demonstrate Understanding that Accommodations Promote Equity in the Class

Students with ADHD can be easily distracted, may be unable to sit still for long periods of time, or may constantly have thoughts racing through their minds. This may cause their test scores to not be reflective of their comprehension of the material. It can also be difficult for students with ADHD to recall information quickly. Accommodations can dramatically promote learning (and thereby increase the test scores) of students with non-apparent disabilities.[28] A fair exam will test an individual's knowledge of required material and not the speed at which they can recall information within a preset time limit. Typically, accommodations, such as extended test time or a reduced distraction environment, are available for students who are formally evaluated and diagnosed with ADHD. Test accommodations may alleviate the anxiety and pressure associated with ADHD and is essential in leveling the "playing field" for all students enrolled and actively pursuing a STEM degree.

It is important for teachers to become familiar with the school's accommodation policy and federal regulations[23-24, 29] so that the process is smooth when a student requests accommodations. Student's willingness to request accommodations is linked to their impression of the instructor's likelihood to treat them with respect (see section 4.1) Some individuals with ADHD are apprehensive about disclosing their diagnoses for fear of judgement or even special treatment.[19] Many students with ADHD want autonomy in their studies and do not want sympathy from the instructors. Empathy is welcome and necessary so that students can feel comfortable discussing accommodations or learning needs.

For students with ADHD (and any diagnosed impairment) are people first, their diagnosis is just a small facet of who they are as students and as people. They want to feel included and valued as members of the class. The most important step an instructor can take may be to extend an open invitation for students to feel comfortable speaking about their needs as learners.